# Commercial Cloud Computing for Connected Vehicle Applications in Transportation Cyber-Physical Systems

**Hsien-Wen Deng**
M. S. Student
School of Computing, Clemson University
351 Fluor Daniel EIB, Clemson, SC 29634
Tel: (864) 643-7646; Email: hsienwd@clemson.edu

**Mizanur Rahman, Ph. D**
Postdoctoral Fellow
Center for Connected Multimodal Mobility ($C^2M^2$)
Glenn Department of Civil Engineering, Clemson University
200 Lowry Hall, Clemson, SC 29634
Tel: (864) 650-2926; Email: mdr@clemson.edu

**Mashrur Chowdhury, Ph.D., P.E., F.ASCE**
Eugene Douglas Mays Professor of Transportation
Glenn Department of Civil Engineering, Clemson University
216 Lowry Hall, Clemson, SC 29634
Tel: (864) 656-3313; Fax: (864) 656-2670
Email: mac@clemson.edu

**M Sabbir Salek***
Ph.D. Student
Glenn Department of Civil Engineering, Clemson University
351 Fluor Daniel EIB, Clemson, SC 29634
Tel: (864) 656-3313; Email: msalek@clemson.edu

**Mitch Shue**
Professor of Practice
School of Computing, Clemson University
226C McAdams Hall, Clemson, SC 29634
Tel: (864)365-0425; Email: mshue@clemson.edu

*Corresponding author

Word count: 5095 words text + 1 table * 250 (each) = 5345 words

**ABSTRACT**

This study focuses on the feasibility of commercial cloud services for connected vehicle (CV) applications in a Transportation Cyber-Physical Systems (TCPS) environment. TCPS implies that CVs, in addition to being connected with each other, communicates with the transportation and computing infrastructure to fulfill application requirements. The motivation of this study is to accelerate commercial cloud-based CV application development by presenting the lessons learned by implementing a CV mobility application using Amazon Web Services (AWS). The feasibility of the cloud-based CV application is assessed at three levels: (i) the development of a cloud-based TCPS architecture, (ii) the deployment of a cloud-based CV application using AWS, and (iii) the evaluation of the cloud-based CV application. We implemented this CV mobility application using a serverless cloud architecture and found that such a cloud-based TCPS environment could meet the permissible delay limits of CV mobility applications. Commercial cloud services, as an integral part of TCPS, could reduce costs associated with establishing and maintaining vast computing infrastructure for supporting CV applications. As the CV penetration levels on the surface transportation systems increase significantly over the next several years, scaling the backend infrastructure to support such applications is a critical issue. This study shows how commercial cloud services could automatically scale the backend infrastructure to meet the rapidly changing demands of real-world CV applications. Through real-world experiments, we demonstrate how commercial cloud services along with serverless cloud architecture could advance the transportation digital infrastructure for supporting connected mobility applications in a TCPS environment.

## INTRODUCTION

Transportation cyber-physical systems (TCPS) integrate cyber systems and physical systems using communication networks and interact with each other to support various applications (*1*)(*2*). In TCPS, transportation applications, which reside in the cyber system, use collected data from the physical system (i.e., different transportation-related sensors, such as connected vehicles (CVs), traffic signals, traffic monitoring cameras) to manage and control traffic on the physical system. TCPS is designed to improve operational efficiency, increase safety, and improve the environmental conditions of the physical system (*1*). To develop a real-time interaction between cyber and physical systems, high-performance computing infrastructure is required to process heterogeneous data from different sources. Here, TCPS implies that CVs, in addition to being connected, communicate with the transportation and computing infrastructure to fulfill application requirements.

Commercial cloud computing services are capable of supporting real-time TCPS applications (*3*)(*4*). For example, Amazon Web Service (AWS), which is a popular cloud service platform, can be utilized to develop cloud-based CV mobility applications in TCPS. Unlike the safety-critical applications such as the forward collision warning, CV mobility applications do not have stringent real-time delay requirements (*3*). For a CV application, the delay refers to a combined delay of the data upload delay from CVs to the TCPS computing infrastructure, the processing time of the application in the computing infrastructure, and the download delay of the feedback generated by the application from the computing infrastructure to the CVs. For mobility applications, which include queue warning and traffic data collection, a delay of up to 1000 ms is acceptable (*5*). Google Cloud Platform (GCP) and Microsoft Azure have similar functionality to AWS to develop cloud-based TCPS applications (*6*)(*7*). These commercial cloud computing platforms are capable of streaming heterogeneous data, securing analysis of data, and preserving privacy. Most importantly, utilizing the available resources of these commercial cloud services platforms, it is possible to dynamically scale-up and scale-down the cloud storage and computing resources depending on the number of vehicles connected to the cloud-based CV applications (*8*).

The motivation for this study is to accelerate cloud-based TCPS deployment by sharing the lessons learned from a commercial cloud-based TCPS, which uses AWS, for CV mobility applications. AWS maintains a vast cloud computing infrastructure and a wide array of services, which makes it highly available, accessible, and scalable for real-time application development (*8*). AWS services include real-time data processing, data transmission, and data archiving capabilities. Studies related to the real-world cloud-based TCPS deployment for CVs are not abundant, however, they are necessary to advance the transportation digital infrastructure research as well as guide public and private agencies on how to implement commercial cloud-based TCPS for CV applications.

Thus, the objective of this study is to identify an effective strategy to utilize AWS for CV mobility applications. In this study, we developed a "vehicle-based traffic surveillance" application using AWS services to evaluate different ways to access and utilize cloud functionality to meet the requirements of the application. We developed the vehicle-based traffic surveillance application using a cloud architecture that features serverless computing rather than traditional servers. Serverless architecture is a new cloud computing execution model now offered by all the leading commercial cloud service providers. The highly automated and decoupling features of a serverless architecture allow developers to focus on the functionality of the CV application and not on installing, configuring, operating, and maintaining traditional server instances (*9*).

In this paper, we present a framework for cloud-based CV applications at three different levels, which includes (i) the development of cloud-based TCPS architecture, (ii) the deployment of the cloud-based TCPS using AWS, and (iii) the evaluation of the cloud-based TCPS for a CV application. In this research, to evaluate cloud-based TCPS, we conducted a case study using the vehicle-based traffic surveillance application to evaluate the feasibility of commercial cloud computing for CV applications.

## RELATED WORK

In (*10*), Herrera-Quintero et al. developed an Intelligent Transportation System (ITS) smart sensor prototype that incorporated commercial cloud services (i.e., AWS and GCP) with a serverless and microservice architecture for a Bus Rapid Transit (BRT) system. By collecting Bluetooth data from BRT users' mobile phones, their sensors can generate the origin-destination (OD) matrix for the bus routes. The authors used GCP Firebase (i.e., Google's mobile development platform) for storing Bluetooth information and AWS Lambda (i.e., an event-driven serverless computing platform within AWS) for calculating the OD matrix. In another study, Karimi et al. introduced a cloud computing-based ITS called "AzureITS" that utilized Microsoft Azure and Ajax (*4*). The AzureITS system can improve traffic conditions and solve the tremendous storage and computing demands for storing and analyzing information related to any roadway crash or injury events observed by the vehicles. Utilizing Azure Service Bus topics (i.e., a publish/subscription-based one-to-many form of communication middleware), AzureITS can help nearby vehicles avoid congestion by providing information related to the crash and any congestion-free alternate routes and can help police and ambulances arrive at the incident location as quickly as possible.

In another study, Howard et al. developed a distributed data analytics framework for smart transportation using AWS-based storage and computing resources (*11*). The authors utilized Amazon Simple Storage Service or S3 (i.e., an object storage service), Amazon Elastic Compute Cloud or EC2 (i.e., a secure and scalable server instance), and Amazon Elastic MapReduce or EMR (i.e., a big data processing and analysis tool) with MongoDB (i.e., a NoSQL database) and open-source Apache Spark (i.e., a distributed cluster-computing facility) for their data analytics framework. Various types of supervised and unsupervised machine learning techniques were evaluated in (*11*) to identify the data sizes for which a distributed system could be useful for machine learning algorithms. Seal and Mukherjee developed a cloud-based real-time accident prediction system for congestion control using Amazon EMR and Apache Spark (*12*). Their cloud-based application uses deep learning for real-time object detection from the streamed video data from connected and automated vehicles. The authors concluded that compared to a single EC2 instance, their EMR-Spark framework can achieve a 60% improvement in terms of cloud-processing performance. In (*13*), Bhatnagar et al. developed an AWS-based car pollution detection warning messaging system. The authors used MQ-7 (i.e., a Carbon Monoxide sensor) for gas leakage sensing and GPS for tracking the car. Using AWS Simple Notification Service or SNS (i.e., a low-cost solution for mass delivery of messages), Bhatnagar et al. were able to send warning messages to the drivers after detecting possible gas leakage.

Commercial cloud services, such as Microsoft Azure, GCP, and AWS, offer serverless computing services. Microsoft introduced "Azure Function" as a part of its serverless computing service offering, which allows users to develop event-driven applications using Visual Studio (*14*). Google Cloud Platform (GCP) offers "Knative" for developers to create and manage their applications using the serverless cloud computing services together with other GCP services, such as cloud storage and data flow (*15*). AWS provides Lambda as its serverless computing service (*16*).

In this study, we utilized a cloud-based serverless architecture to support our CV mobility application using AWS. The same approach can be used to develop serverless CV applications using Microsoft Azure or GCP. AWS maintains a vast cloud infrastructure and a wide array of services, which makes it well-suited for real-time CV application development. Additionally, as shown in (*3*), a comparison of the serverless computing platforms offered by AWS, GCP, and Microsoft Azure in terms of processing time revealed that AWS Lambda can provide the best performance in terms of processing time. Therefore, we chose AWS as the cloud service provider and AWS Lambda as the event-driven serverless computing service in this study.

## CLOUD-BASED TRANSPORTATION CYBER-PHYSICAL SYSTEM FOR CVs

A cloud-based TCPS introduce features of commercial cloud computing and integrates commercial cloud services with other TCPS components. The wide array of services offered by commercial cloud service providers encourages the development and implementation of CV applications in a TCPS environment. A commercial cloud can be very cost-efficient for TCPS applications in terms of hardware installation and maintenance. **Figure 1** presents a general architectural approach for a cloud-based TCPS. CVs, connected vulnerable road users (e.g., pedestrians, cyclists) and transportation infrastructure (e.g., traffic management infrastructure, traffic control infrastructure, roadside data infrastructure) can send their data directly to the cloud-based application and receive corresponding application output. The cloud infrastructure contains three layers: (i) infrastructure, (ii) platform, and (iii) software. The infrastructure layer, which is also known as infrastructure-as-a-service (IaaS), provides hardware components to the application developers (or cloud users) to set up virtual machines (i.e., server instances), install operating systems and provide database resources. The platform layer, which can be defined as platform-as-a-service (PaaS), allows application developers to build an application using various cloud services, such as database management, computing, streaming and non-streaming services without being concerned with managing the underlying infrastructure. Application developers can use the software layer (i.e., software-as-a-service (SaaS) layer) to deploy their TCPS application in the cloud using the uploaded data from different transportation data sources including CVs and receive output from the application running in the cloud.

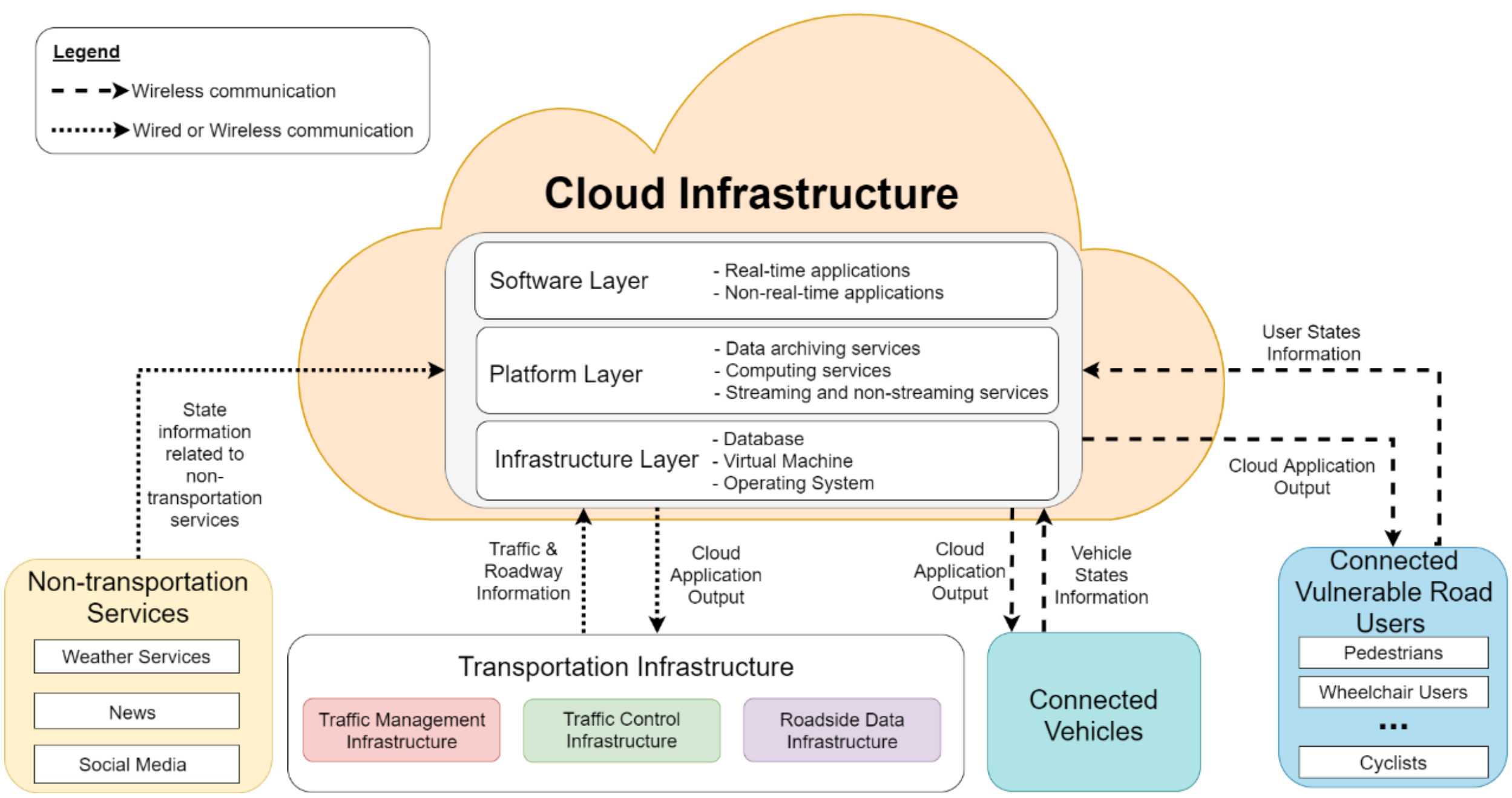

**Figure 1 Cloud-based transportation cyber-physical system for CVs**

A cloud-based TCPS application can be real-time or non-real-time depending on the CV application requirements. To enable cloud-based TCPS, cloud services must fulfill the quality of service (QoS) agreements in terms of computing and communication delay requirements. In this study, we provide a systematic view of a cloud-based TCPS for CVs using several cloud services for computing, data transmission, and data archiving. The following subsections present the details of different components of the cloud-based TCPS for CVs: (i) real-time cloud computing; (ii) real-time data transmission; and (iii) data archiving.

## Real-time Cloud Computing

A cloud computing service provides the computing capacity for a cloud-based TCPS application and is responsible for processing data after collecting it from databases and other cloud services. Traditionally, when using a server-based architecture, application developers need to establish a cloud server instance according to their computing requirements. This requires significant time and expertise for hardware set up and virtual environment configuration, and it does not allow for the dynamic scaling of computing resources based on the demand. Recently, a new cloud computing architecture, known as serverless architecture, allows application developers to focus on application functionality and relieves them from the burden of setting up or configuring the hardware and virtual environment. In a serverless architecture, the cloud computing environment can dynamically allocate the necessary compute resources to the application without developer intervention. Thus, serverless architecture can provide a cost-effective alternative to establishing and maintaining server instances. The following sub-subsections present the details of the server-based and serverless cloud computing.

*Server-based Cloud Computing*

A server instance is the computing engine of a server-based TCPS cloud application (e.g., real-time or non-real-time). As **Figure 2** illustrates, application developers first establish a server instance in the commercial cloud. This requires configuration of virtualized or dedicated hardware (e.g., CPU, memory, storage) and software environments, such as operating systems (OS) as well as coding platforms (e.g., language environment, compilers, libraries) for a given application. The application interacts with other cloud services through various application programming interfaces (API) to complete the functionality of the application. A server-based cloud application allows developers to customize computing capabilities based on demand. This approach, however, requires developers to spend more time and effort on configuring and maintaining server instances. Moreover, computing capacity in a server instance is fixed and can be wasted if the application requires fewer resources.

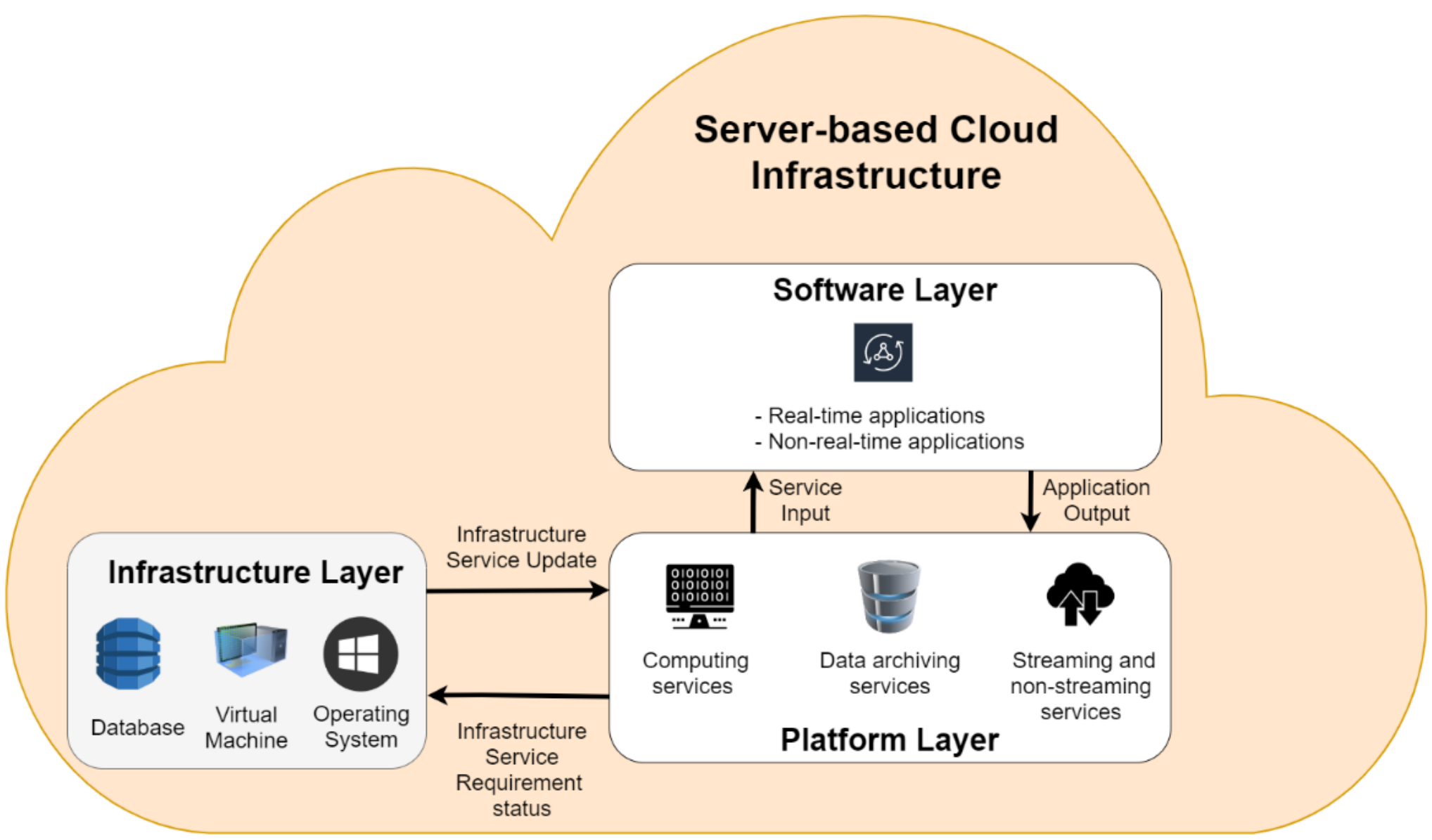

**Figure 2 Server-based computing concept**

*Serverless Cloud Computing*

Commercial cloud service providers offer a new cloud computing architecture that does not require developers to establish server instances. This is known as serverless architecture (as shown in **Figure 3**). Without the burden of creating, configuring, maintaining, and operating server instances, developers can focus on developing the functionality of the TCPS application. In general, serverless applications are event-based and can be triggered by the actions of other cloud services, such as database update service and the arrival of input data. The serverless function supporting the application launches automatically after being triggered and processes until it completes. The serverless compute service supports various programming languages, including Python, .NET, Java, and Node.JS. The process is highly automated and once a serverless function finishes its job, the commercial cloud releases the computing resources used. From an application developer perspective, this approach saves time, effort, and costs compared to traditional server-based applications. Serverless architectures are capable of employing computing resources more efficiently.

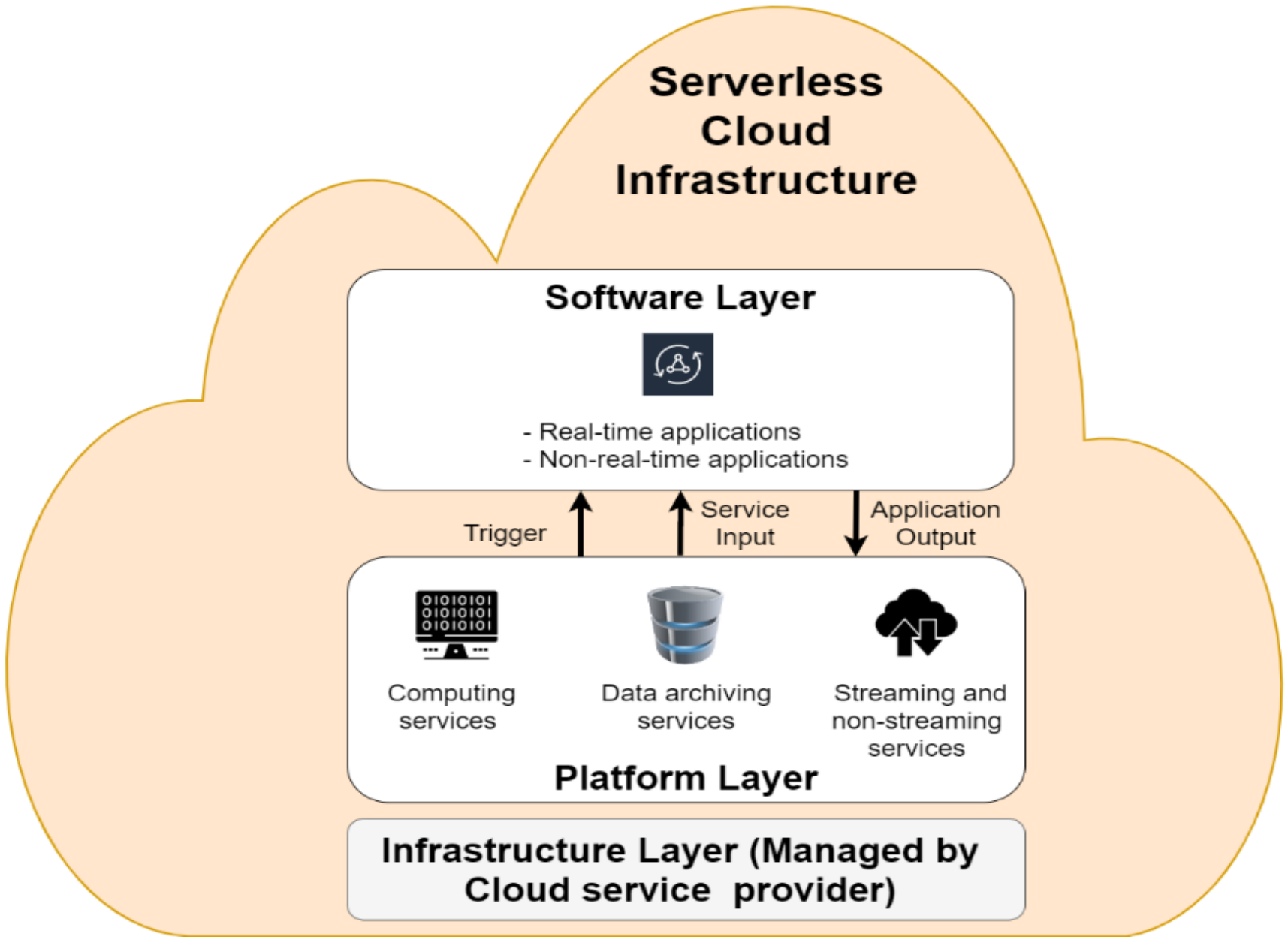

**Figure 3 Serverless computing concept**

**Real-time Data Transmission**

With the Internet of Things (IoT), all connected physical devices in a TCPS are treated as connected nodes. Transportation infrastructure, such as roadside data infrastructure and traffic control infrastructure (e.g., traffic signal) can be recognized as fixed edge nodes, whereas CVs and connected vulnerable road users can be recognized as mobile nodes. Based on different characteristics of roadside infrastructure and vehicles, they can use either wired or wireless communication to transmit their data to the cloud. As **Figure 1** illustrates, transportation infrastructure uses either wired connections (e.g., optical fiber) or wireless connections (e.g., Long-term evolution (LTE) and 5G) to communicate. On the other hand, as mobile nodes, CVs and connected vulnerable road users can only transmit data to the cloud through wireless communication (e.g., LTE and 5G). An on-board unit (OBU) in a CV or a cell phone of a vulnerable road user can also upload or download relevant information directly to or from the cloud through a wireless communication network, such as LTE and 5G.

**Real-time Data Archiving**

For storing historical data, cloud database services can be used to archive data collected from CVs, connected vulnerable road users, transportation infrastructure, and other TCPS services, such as weather, social media, and news services in a TCPS environment. These historical data can be used by any cloud-based CV application based on the application requirements. Moreover, to support real-time CV applications, databases need to share data with other cloud services, such as cloud computing service, with a low communication delay to fulfill required QoS agreements.

**CASE STUDY**

We conducted a case study to evaluate the feasibility of cloud computing for TCPS in the context of a real-time CV mobility application, i.e., vehicle-based traffic surveillance. We developed the application utilizing serverless architecture in AWS rather than server-based architecture. Experiments from a prior study revealed that serverless computing in the cloud using AWS Lambda yielded the best performance in terms of processing time among the serverless architecture provided by other commercial cloud service providers (*3*). We also analyzed the availability and reliability of the cloud during various times of the day and considered the scalability required of the cloud services based on the demand for accommodating dynamic traffic conditions.

**Vehicle-Based Traffic Surveillance Application**

In this study, we deployed a vehicle-based traffic surveillance application in the serverless cloud. In our experiment, CVs on the roadway transmit data to a database service in the cloud (e.g., NoSQL database) through the LTE communication network in real-time and trigger the vehicle-based traffic surveillance application in the cloud. The application then collects vehicle data (e.g., speed and location) from the CVs and processes them. After processing, data are stored in a database as cloud application feedback, which is the average speed on the roadway. CVs receive this feedback by subscribing to the application, which informs CVs of the average speed of their traveled roadway section.

**Serverless Cloud-based TCPS Architecture**

In AWS, we implemented a vehicle-based traffic surveillance application in a serverless cloud-based TCPS to collect, aggregate, and process the Basic Safety Messages (BSMs) from CVs. DynamoDB is a NoSQL database based on key-value pairs, which we used as data storage in the cloud (*17*). AWS Lambda is the serverless computing service we used to build the application (*16*). As shown in **Figure 4**, two DynamoDB databases are established: (i) vehicle trajectory database and (ii) feedback database. CVs send BSMs to the vehicle trajectory database and then trigger the vehicle-based traffic surveillance application, which runs as an AWS Lambda function. The application utilizes vehicle trajectory information (e.g., BSMs) from the vehicle trajectory database and processes them. After completion, the Lambda function stores traffic surveillance results in the feedback database. Finally, CVs subscribe to the feedback database to receive the average speed on their traveled roadway. In this study, we developed our application using AWS services (e.g., DynamoDB, AWS Lambda) all of which are located in the us-east-1 AWS service region (N. Virginia) because of its proximity to our location.

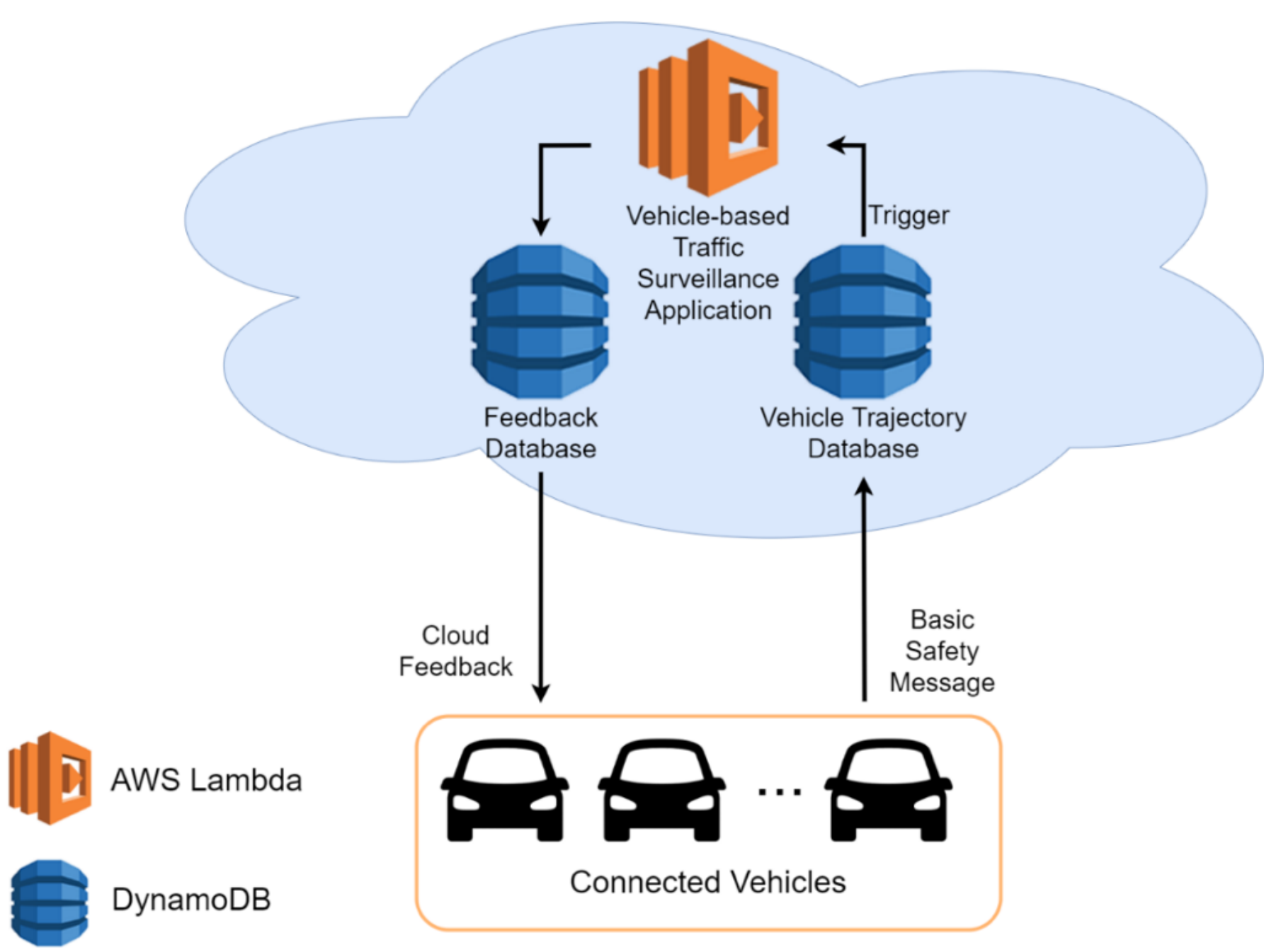

**Figure 4 Vehicle-based traffic surveillance application in AWS**

**Experimental Setup**

We conducted our experiments using three CVs along Perimeter Road on the main campus of Clemson University in Clemson, South Carolina. It is known as the South Carolina Connected Vehicle Testbed (SC-CVT), with a true TCPS environment. In **Figure 5**, Perimeter Road (highlighted in the purple color) indicates trajectories of CVs, and the arrows in the orange color indicate the direction of traffic flow. The speed limit of the roadway is 35 mph and all three vehicles were operated within the speed limit while we conducted our field experiments. In each vehicle, we ran a Python script that automatically measured communication delays by sending BSMs to the cloud and receiving average speed information from the cloud after running the vehicle-based traffic surveillance application in the cloud. The transmissions of data between CVs and AWS are all processed through LTE. We collected 1000 samples for uploading data to the cloud and 1000 samples for downloading data from the cloud for each CV. We also collected processing time to measure the performance of our serverless application, separate from any network delay. Finally, we analyzed the processing time, the upload communication delay, the and download communication delay to evaluate the feasibility of this cloud-based CV mobility application in terms of QoS related to delay (i.e., the delay associated with communicating with the cloud and processing time in the cloud).

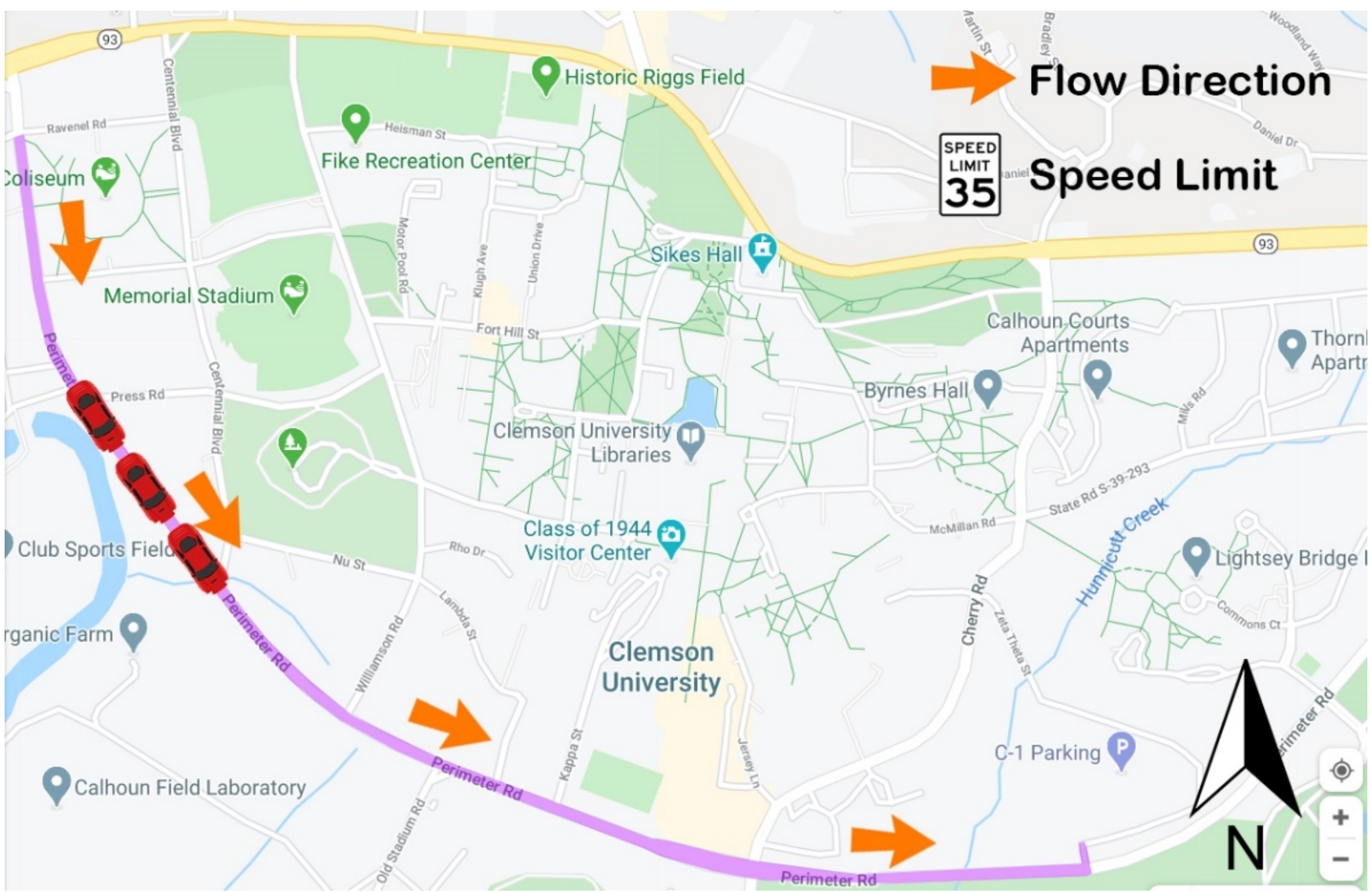

**Figure 5 Field experiment configuration**

**Field Test Results and Discussion of Cloud-based TCPS Feasibility**

We calculated the average and 95$^{th}$ percentile of the upload and download communication delays associated with AWS for three vehicles and the application processing time in the cloud. The 95$^{th}$ percentile value indicates a threshold QoS guarantee (*18*)(*19*). **Figure 6** shows that both 95$^{th}$ percentiles of upload delay and download delay were less than 140 ms. In the serverless cloud-based TCPS, the vehicle-based traffic surveillance application took only 74 ms on the 95$^{th}$ percentile scale to process the data (as shown in **Figure 7)**. As shown in **TABLE 1,** we observed an average upload delay of 85 ms and an average download delay of 84 ms, with an average processing time of 41 ms in the serverless cloud. On the 95$^{th}$ percentile scale, we observed the upload delay as 136 ms and download delay as 139 ms, and 74 ms as the processing time in the cloud. This means that performance for an individual CV stayed within the QoS requirements as the 95$^{th}$ percentile of total delay - which includes round-trip time (RTT) or end-to-end delay and processing time for vehicle-based traffic surveillance, is found to be only 349 ms, which is lower than the application requirement of 1000 ms (*5*).

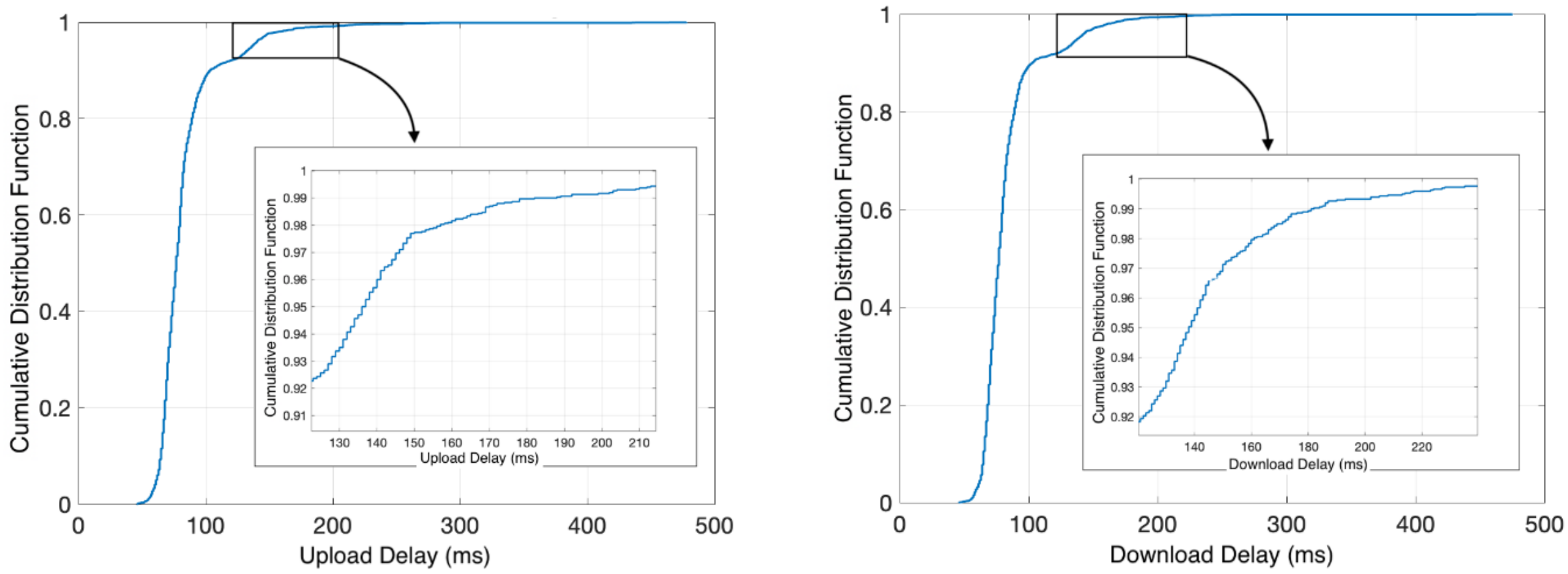

**Figure 6 Cumulative distribution function of upload and download delays**

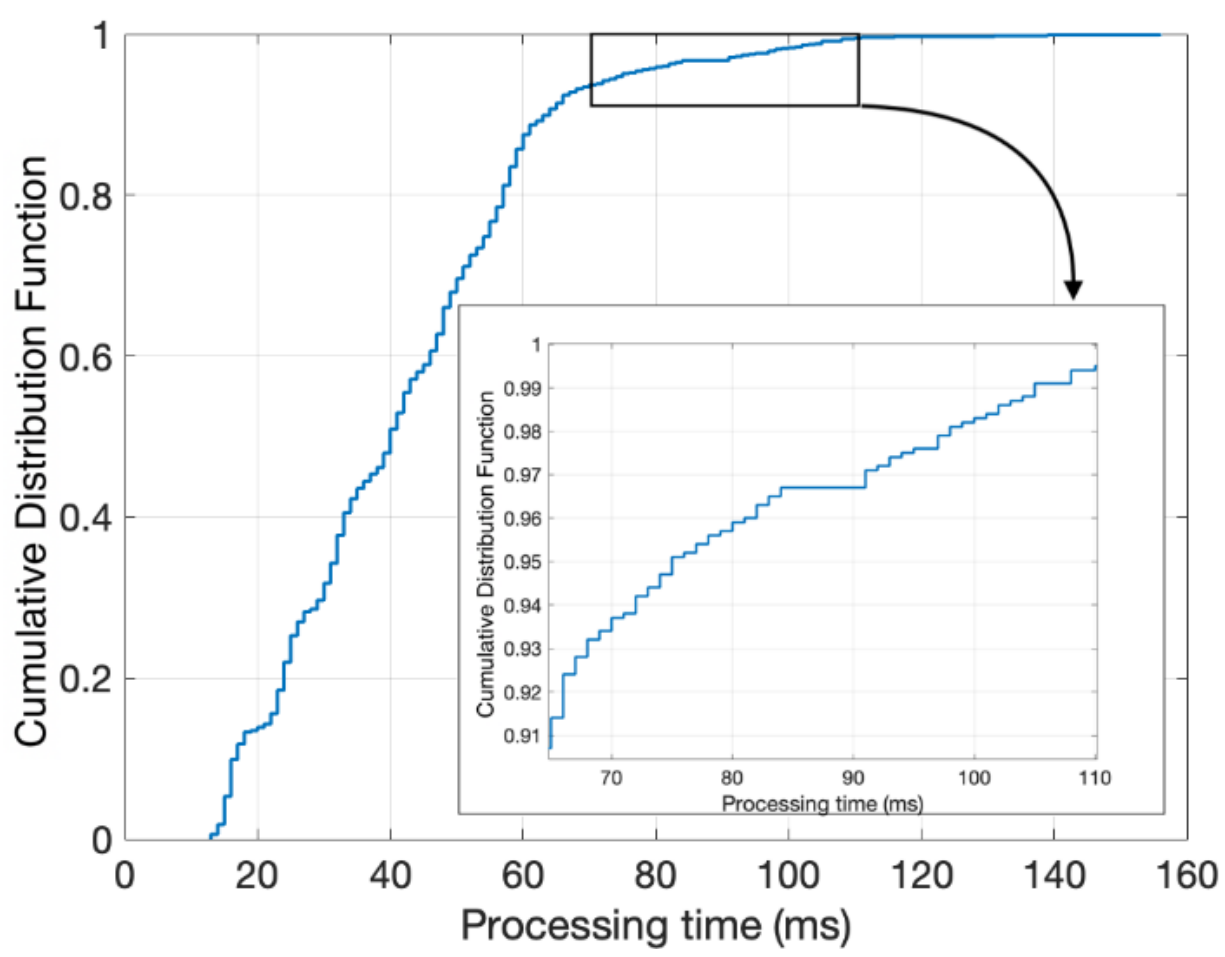

**Figure 7 Cumulative distribution function of application processing time in Lambda**

**TABLE 1 Average and 95th Percentile of Communication Delay**

| | Average | 95th Percentile | Allowable Delay (ms) |
|---|---|---|---|
| Upload Delay (ms) | 85 | 136 | - |
| Download Delay (ms) | 84 | 139 | - |
| Process Time (ms) | 41 | 74 | - |
| End-to-end Delay (ms) | 210 | 349 | <1000 |

**Availability and Scalability of Commercial Cloud-based TCPS**

*Availability*

Due to dynamic changes in LTE communication bandwidth usage at different time frames of a day, communication delays may vary. We experimented with our vehicle-based traffic surveillance application in three-hour granularities for 24 hours of a day (i.e., once every three hours) to estimate the communication delay over one day. For each experiment, we collected 1000 samples on both upload and download delays. Then, we determined the 95$^{th}$ percentile of the samples to ensure that delay for cloud access to support whether the CV application can be evaluated against QoS requirements (*18*)(*19*). **Figure 8** illustrates over one day that the 95$^{th}$ percentile of upload and download delays did not exceed 100 ms. We also identified a relative peak hour for communication bandwidth usage. As **Figure 8** shows, 12:00 PM - 3:00 PM (15:00) timeframe is observed as the peak hours for uploading data. For downloading data, 03:00 AM - 09:00 AM, 12:00 PM - 3:00 PM (15:00) and 06:00 PM (18:00) - 09:00 PM (21:00) can be considered as peak hours based on our experiments. These periods were based on our local time zone which is same as the us-east-1 AWS service region (N. Virginia).

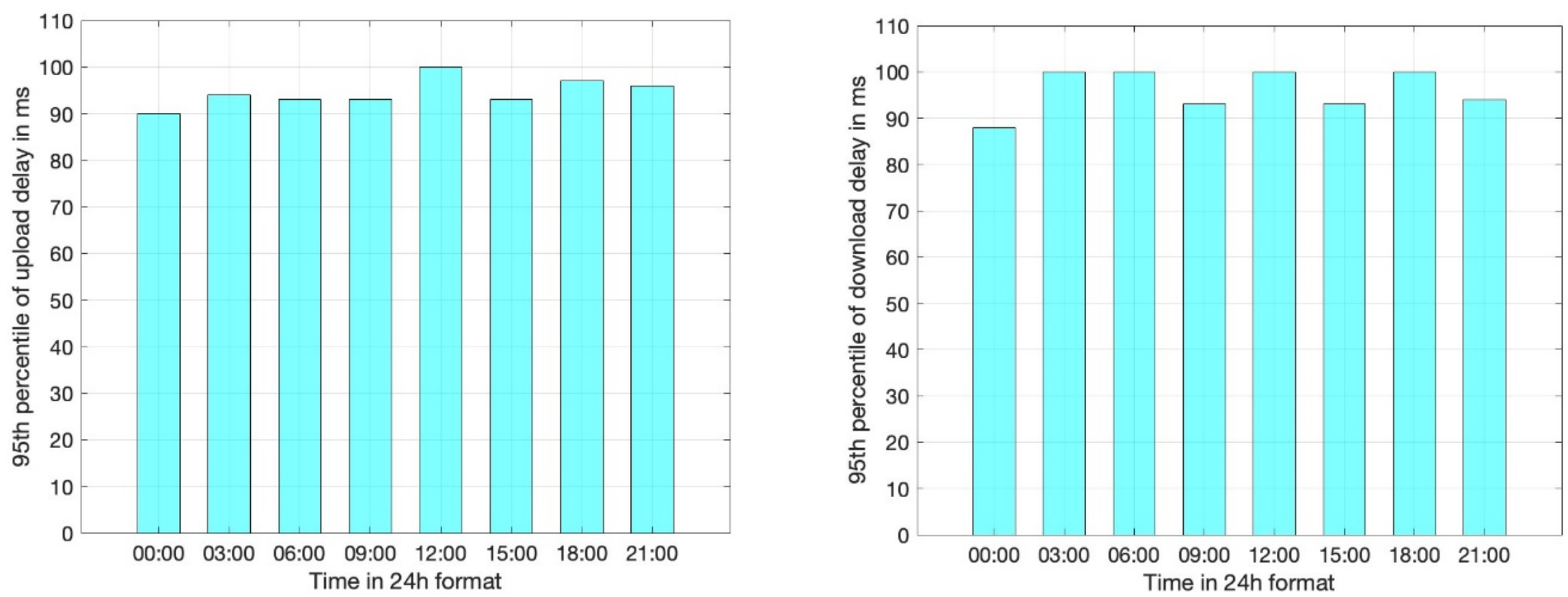

**Figure 8 Cloud accessibility in terms of 95$^{th}$ percentile of communication delay**

*Scalability*

Due to the dynamic scaling characteristics of traffic demand, a vehicle-based traffic surveillance application in the cloud must also scale automatically. Although serverless architecture supports dynamic resource allocation, there are a few limitations, which are specified by the commercial cloud service providers. For example, AWS currently allows a single Lambda function to utilize up to 3008 MB of memory when processing. Because of this, a single Lambda function may not be able to support large scale TCPS in a real-time manner. To address this issue, we applied the concept of parallel computing in the cloud. As **Figure 9** shows, in our vehicle-based traffic surveillance application, we established a group of Lambda functions to handle changing traffic conditions in parallel at scale. We can define a computing capacity for every Lambda function (i.e., the number of vehicles to process). All Lambda functions operate in parallel to improve throughput. Using this parallel approach, the mobility application can process a large amount of data as traffic demand increases without exceeding the maximum allowable latency.

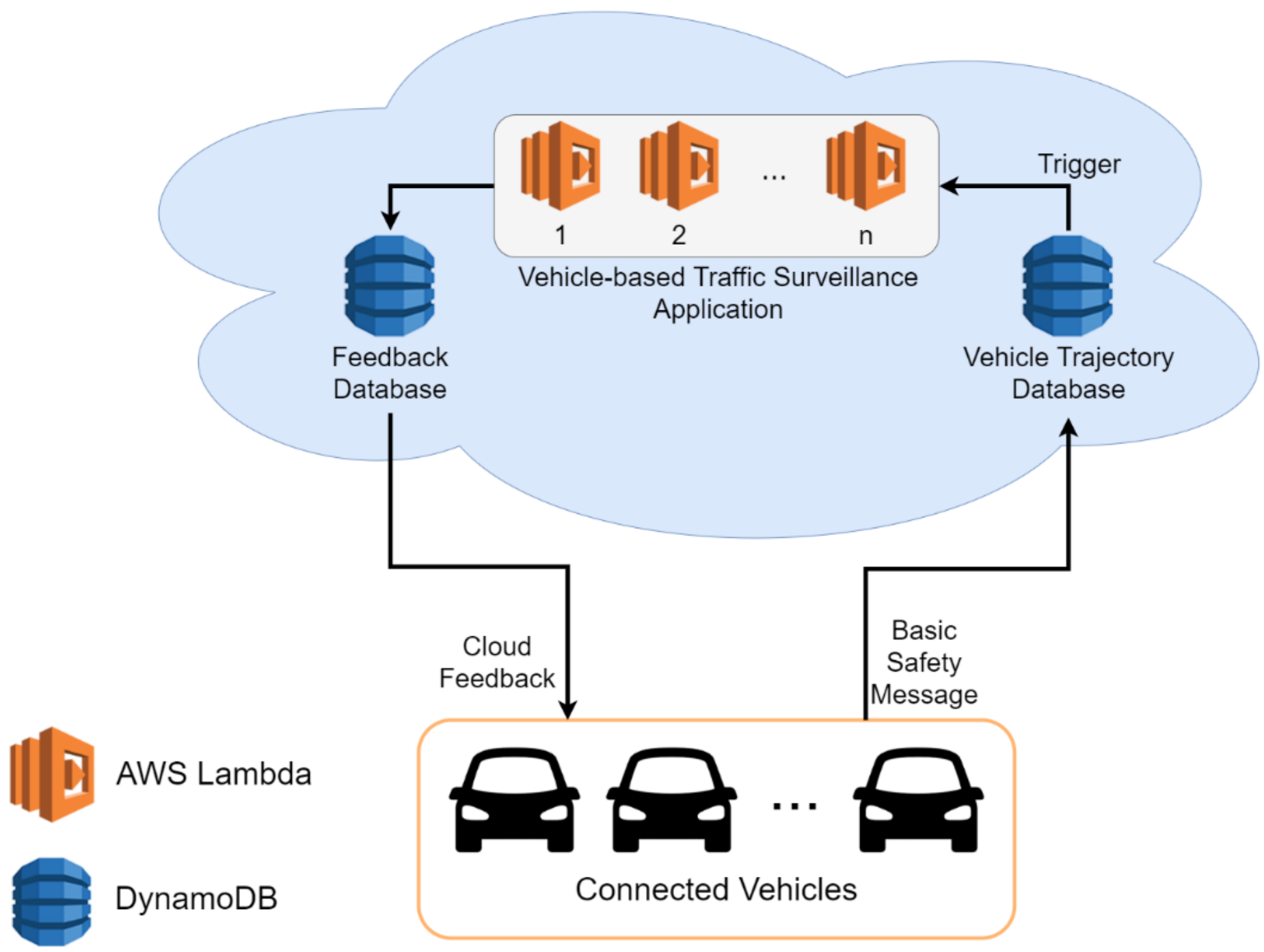

**Figure 9 Parallel computing method in AWS serverless architecture**

## CONCLUSIONS

As public agencies are responsible for operating and managing the transportation systems, cloud computing could be a potential alternative to significant investments in equipment and human resources to support a local computing infrastructure for CVs in a TCPS environment. Significant backend computing infrastructure is needed to support the processing of data from CVs and other related data to support CV applications in a TCPS environment. Thus, commercial cloud computing services could become a more viable option for public agencies to develop TCPS for accommodating different levels of penetrations of connected vehicles in the coming years. CVs will increasingly become a part of the mainstream transportation system and commercial cloud computing services could be the solution for public agencies to support mobility applications related to CVs. Commercial cloud infrastructure offers promising functionality for addressing scaling issues related to traditional computing infrastructure as more and more CVs are added to our transportation system in the coming years.

The case study presented in this paper argues for a vehicle-based traffic surveillance application through AWS services. We demonstrated the usage of commercial serverless cloud architecture for processing a real-time mobility application for CVs. This study revealed feasibility of developing a CV application using a serverless cloud architecture that can meet the delay requirements of CV applications, without the burden of provisioning, configuring, operating, and maintaining server instances. Moreover, serverless cloud architecture is more cost-efficient and provides the scalability and flexibility needed for robust CV applications.

Our analysis revealed that, on average, RTT between CVs and the cloud was 210 ms in which processing time within the cloud was only 41 ms. These results show that commercial cloud service providers offer a reliable and secure capacity to support robust CV mobility applications. Bringing TCPS applications to market using serverless cloud architecture can meet the maximum allowable delay requirements for traffic mobility applications and also reduces the cost to establish and maintain computing infrastructure (*20*).

## FUTURE RESEARCH FOR CLOUD-BASED TCPS

Future research should focus on comparing benefit-cost analysis of using the commercial cloud over public agencies' local computing infrastructure. This analysis will provide important guidance on the best use of public investment for computing infrastructure by agencies responsible for supporting CV operations on public roads.

Future research should evaluate the relative performance of the serverless and server-based cloud architecture for CV applications in a real-world testbed with different penetration levels of CVs. Evaluation parameters should include cost, delay, and reliability associated with different CV applications.

Future research should compare different commercial cloud service providers, such as Microsoft Azure, GCP, and AWS for supporting different CV applications in the real world.

Commercial cloud services that can support CV applications must be secure to prevent CV applications from being compromised. Future research should identify cybersecurity risks associated with CV applications using commercial cloud services and various defense-in-depth strategies to protect these applications. Critical security elements, which include identity, confidentiality, integrity, availability, authentication, accountability, and privacy, must be evaluated for cloud services accessed by CVs.

Although this study employed AWS cloud infrastructure to support our real-time CV mobility applications, it is important to evaluate the performance and reliability of other commercial cloud service providers for CV safety applications (e.g., collision avoidance) due to the low communication latency requirements for safety applications like these.

The availability of on-demand computing services and the reliability of cloud services make them suitable for developing CV applications with an increasing number of CVs. Future research should focus on the scalability and high availability characteristics of cloud computing for applications handling increased penetration levels of CVs on roadways.

## ACKNOWLEDGMENTS

This study is based on a study supported by the Center for Connected Multimodal Mobility ($C^2M^2$) (USDOT Tier 1 University Transportation Center) Grant headquartered at Clemson University, Clemson, South Carolina, USA. Any opinions, findings, and conclusions or recommendations expressed in this material are those of the author(s) and do not necessarily reflect the views of the Center for Connected Multimodal Mobility ($C^2M^2$), and the U.S. Government assumes no liability for the contents or use thereof.

## AUTHOR CONTRIBUTIONS

The authors confirm contribution to the paper as follows: study conception and design: H. Deng, M. Rahman, M. Chowdhury, M. Salek, and M. Shue; data collection: H. Deng, M. Rahman and M. Salek; interpretation of results: H. Deng, M. Rahman, and M. Salek; draft manuscript preparation: H. Deng, M. Rahman, M. Chowdhury, M. Salek, and M. Shue. All authors reviewed the results and approved the final version of the manuscript.